# A Proposal To Support Wellbeing in People With Borderline Personality Disorder: Applying Reminiscent Theory in a Mobile App

Alice Good, Clare Wilson, Claire Ancient and Arunasalam Sambhanthan
University of Portsmouth
Portsmouth, Hampshire, UK
alice.good@port.ac.uk

**ABSTRACT**
In this paper the research draws upon reminiscence therapy, which is used in treating dementia, as an applied theory to promote well being in people who experience low moods. The application proposed here aims to promote wellbeing for people suffering from mood disorders and dementia but could potentially be used to enhance wellbeing for many types of users. Use of the application is anticipated to improve mood in a group of users where severe emotional problems are prevalent. The research aims to evaluate the effectiveness of a reminiscence based application in promoting well being in people specifically with Borderline Personality Disorder (BPD). The long term objective of this research is to establish the effectiveness of reminiscence theory on user groups aside from dementia, particularly other mental illnesses. The research advocates involving end users within the design process both to inform and evaluate the development of a mobile and tablet application.

**ACM Classification Keywords**
HCI, health

## INTRODUCTION
The fourth version of Diagnosis & Statistical Manual of Mental Disorder classifies Borderline Personality Disorder as a mental disorder which is characterized by a range of debilitating and self destructive behaviours including: depression [1]. People with BPD in the UK are reported as more likely to seek intervention than people with other psychiatric disorders and in fact 4-6 % of patients receiving primary care have been diagnosed with BPD [3]. Treatment for people BPD has traditionally included both pharmacological and psychological intervention. These methods have their issues including the fact that the latter may not be practical or timely. Often, the identified user group struggle alone with managing their daily life and moods. Furthermore, traditional methods are often conducted in a patriarchal manner and are disempowering to the service user. Ubiquitous methods of enabling these people to better support their well being could certainly be useful to them, as well as enabling them to ultimately have control over their own lives.

## USER-CENTRED DESIGN IN DEVELOPING APPS
In investigating how IT applications can better support people with BPD, the researchers are currently investigating the level of support available, whether this meets requirements and what other forms of help would be useful. This has been carried out from both a service provide and user perspective. Preliminary results indicated a strong need for accessible and timely support, which for most people with BPD was not available [2]. Whilst the aim of the research was to investigate the need for support, the interviews with service users with BPD indicated a strong willingness to engage in any research that might be required in designing supportive ICT applications. It is of course well documented that involving end users within the design process lends to a stronger sense of ownership [5].

### Self Soothing
The act of 'self soothing' that is calming us down is one of the hardest things to do when you have mental health problems. Yet the ability to be able to calm oneself down, to essentially self soothe would be advantageous to people with BPD, and could effectively prevent a potential crisis from escalating. A mobile application that could bring about a sense of positivity and achievement could also help train the people to self soothe themselves, so that ultimately, they would no longer require the app.

There is a steady growth in the development of mobile apps that aim to help people manage their lives. Predominantly these are designed without a theoretical research basis and are not necessarily user centred in their approach to design. Where apps are designed to support well being, to manage lives and to 'self soothe', there is then an onus not only upon the app fundamentally matching the user's goals but also on providing a positive user experience.

**REMINISCENT THEORY IN DEMENTIA TREATMENT**

Recent statistics from the Alzheimer's society state that there will be approximately one million people diagnosed with Dementia in the UK by 2021 [6]. This debilitating syndrome strips people of their precious memories which have accumulated over time. Reminiscent therapy (RT) however, is a popular method used in promoting positive mood and well being and reducing the sense of feeling alone for people with dementia (See Figure 1). It involves using meaningful prompts, including photos, music and recordings as an aid to remembering life events [4]. Some research states that it has been useful in reducing depression [7]. Whilst it has been predominantly utilized in people with dementia, there could be scope for applying this theory in other mental health conditions where depression and general low mood are common. This could potentially induce a 'self soothing' process which could lend itself well to people who struggle with day to day living as a result of low mood.

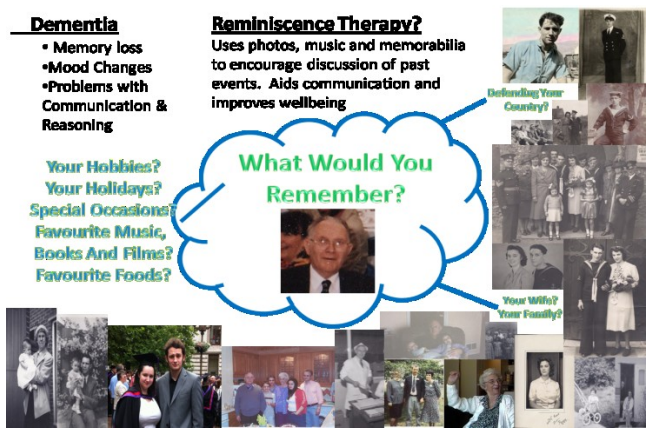

**Figure 1. Reminiscent Therapy for Dementia**

Given the prevalence of low moods and a poor sense of wellbeing in people with BPD, RT could be applied to this user group. Furthermore, people with BPD often experience a sense of feeling alone and 'spiraling' negative thinking. To this extent, the researchers have selected to use this user group to assess the effectiveness of an application based upon RT.

**MY FAVOURITE THINGS**

'My Favourite Things' is intended to be an application that is designed upon the theory of RT. Users can upload meaningful memorabilia including: photos; videos; songs; recordings and notes. The collated memorabilia can then be viewed in various formats. One example could be a slide show of photos with music attached. Another example could be a categorization of selected items that the user could choose to view from (See figure 2). The application can be set up by individuals or with the support of service providers, for example community nurses/support workers.

The application is not intended to replace therapy but to serve as a standalone support mechanism. Based upon RT theory, the application could potentially promote positive mood and therefore increase a sense of well being.

In adopting a user-centered approach to design, the researchers intend to conduct a focus group, comprised of people with BPD, to inform and evaluate the design of a prototype mobile application.

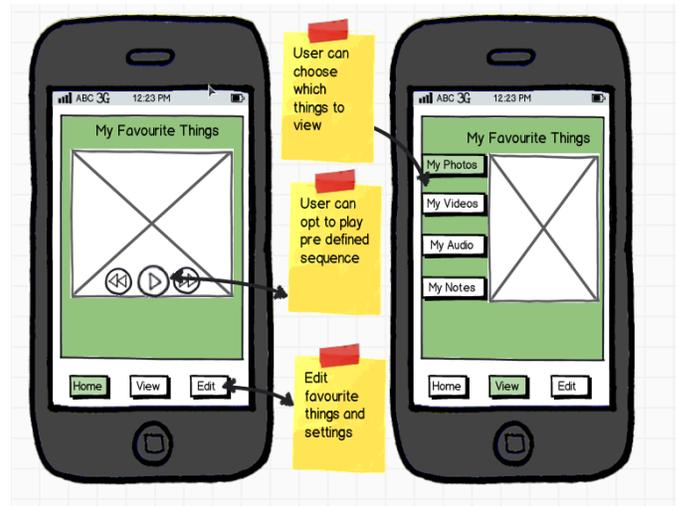

**Figure 2. Prototype for My Favorite Things**

Subsequent implementation of the application will lead to evaluating the effectiveness of these apps on users with BPD. Further research will then explore the effectiveness of 'My Favourite Things' App in other user groups.